\RequirePackage{fix-cm}
\documentclass[twocolumn,epjc3]{svjour3}
\smartqed  
\RequirePackage{graphicx}
\RequirePackage{mathptmx}      
\RequirePackage{latexsym}
\RequirePackage[numbers,sort&compress]{natbib}

\RequirePackage[colorlinks,citecolor=blue,urlcolor=blue,linkcolor=blue]{hyperref}
\journalname{Eur. Phys. J. C}
\begin{document}

\title{Comment on the  paper "Generalized version of chiral Schwinger model in
terms of chiral bosonization", Eur.Phys.J.C 81 (2021) 3, 199.
} 


\author{Anisur Rahaman\thanksref{e1,addr1}
}

\thankstext{e1}{e-mail: manisurn@gmail.com, anisur.rahman@saha.ac.in}


\institute{Durgapur Government College, Durgapur-713214, Burdwan
West Bengal, India \label{addr1}
}

\date{Received: date / Accepted: date}

\maketitle

\begin{abstract}
The results and calculations of the article \cite{SY} are found
erroneous for a generic $r$ and it remains irremediable in
general. However for $r=0$, $r=\pm 1$, and $r >> 1$ it may be
physically acceptable.
\end{abstract}

The author of the article \cite{SY}, recently carried out an
investigation to find out the theoretical spectra of a model which
was resulted after imposition of a chiral constraint by Miao
\cite{MIA}, in the phase-space of the generalized chiral Schwinger
model (GCSM) \cite{BASS, WOT, SAR}. The article \cite{MIA} however
was not cited in \cite{SY}. The innovative idea of imposing chiral
constraint was put forward by Harada in his seminal work \cite{KH}
where he imposed a chiral constraint in the Chiral Schwinger model
(CSM) and expresed it in terms of chiral boson \cite{SIG, JSON}.
The generating functional of the GCSM is given by
\begin{eqnarray}
Z(A)=\int{d\psi} {d\bar{\psi}}e^{i
\int{d^2x}\bar{\psi}\gamma^{\mu}[i\partial_{\mu}+e\sqrt{\pi}A_{\mu}(1-r\gamma_{5})]}
.\label{INT}
\end{eqnarray}
The bosonized version \cite{BASS, SAR, MIA} corresponding to the
fermionic Lagrangian density  constituting Eqn. (\ref{INT}) reads
\begin{eqnarray}
{\cal L}_B=\frac{1}{2}
\partial_{\mu}\phi\partial^{\mu}\phi+eA_{\mu}(\tilde{\partial}_{\mu}+r\partial_{\mu})\phi
+ \frac{1}{2}ae^2A_\mu A^\mu,\label{LAGB}
\end{eqnarray}
The GCSM contains both the vector and axial-vector interaction
terms with the unequal wight. For both the choices $r=\pm 1$ the
GCSM corresponds to CSM \cite{JR, RB, MIT, ROT1, ROT2}. Fermion of
left/ right chirality takes part in interaction with the gauge
field for $r= +1$/$r= -1$ respectively. Here
$\tilde{\partial}_{\mu}= \epsilon_{\mu\nu}\partial^\nu$ and
$\epsilon^{01} =+1$ and $\psi, \phi$, and $A_\mu$ respectively
represent fermion, boson, and gauge field . The Lagrangian
(\ref{LAGB}) after the  imposition of the chiral constraint
$\Omega=\pi_\phi-\phi'=0$, along with the kinetic term of the
electromagnetic background $\frac{1}{2}(\dot{A_1}-A_0')^2$ turns
into
\begin{eqnarray}
 L_{CB}&=&
\dot\phi\phi' -\phi'^2 +(1+r)(A_0-A_1)\phi'  +
\frac{1}{2}(\dot{A_1}-A_0')^2
\nonumber \\
&-& \frac{1}{2}e^2(A_1-rA_0)^2 + \frac{ae^{2}}{2}(A_0^2 -A_1^2).
\label{LAGCH}
\end{eqnarray}
 The Lagrangian and Hamiltonian
formulation of the model (\ref{LAGCH}) is attempted in \cite{SY}
to study the theoretical spectrum of the model. The author claimed
that the model was exactly solvable for the generic $r$ and
theoretical spectrum contained {\it only a massive boson} with the
square of the mass $m^2= ae^2\frac{r^2+a-1}{a-r^2}$ which showed a
sharp disagreement with the result found in \cite{KH}. For
instance setting $r=1$ in \cite{SY} although a massive boson was
found the massless chiral boson was found absent. This absence of
massless chiral boson in \cite{SY} stood as a grave disparity so
far counting of  number of  physical degrees of freedom is
consented. Landing onto this erroneous result the author claimed
that the fermion got confined \cite{SY}. How the article \cite{SY}
is suffering from erroneous result and to what extent it is
remediable that we would like to present through this note.

To determine the theoretical spectrum transparently through
Hamiltonian analysis of the Lagrangian density (\ref{LAGCH} we
compute the momenta corresponding to the field $\phi$, $A_{0}$,
and $A_{1}$:
\begin{equation}
\pi_{\phi}=\dot{\phi}-eA_{1}+erA_{0},~~
\pi_{\phi}=\phi',~~~\pi_{0}=0,
~~\pi_{1}=\dot{A_{1}}-A_{0}^{\prime}. \label{MOMA1}
\end{equation}
Note that $\Omega_1=\pi_{0}\approx 0$, is a primary constraint of
the theory. The canonical Hamiltonian density is now obtained
using Eqns. (\ref{MOMA1}) by exploiting a Legendre transformation:
\begin{eqnarray}
{\cal H}_{c}&=&\frac{1}{2}\pi_{1}^{2} + \phi'^{2}
+\pi_{1}A_{0}^{\prime}+\frac{1}{2}e^{2}(A_{1}-rA_{0})^{2}\nonumber\\
&-& e(1+r)(A_{1}-A_{0})\phi^{\prime}
-\frac{ae^{2}}{2}(A_0^2-A_1^2). \label{CHAM1}
\end{eqnarray}
The preservation of $\Omega_1$ results a secondary constraint
\begin{equation}
\Omega_{2}=\pi'_1+e^2((a-r^2)A_{0}+rA_1)+e(r+1)\phi'\approx 0.
\label{CON2}
\end{equation}
 $\Omega_1$ and $\Omega_1$ are weak conditions \cite{DIR} and form a second
class set. From Eqn. (\ref{CON2}) we have a solution for $A_{0}$:
\begin{equation}
A_{0}=-\frac{1}{e^{2}(\alpha-r^{2})}(\pi_{1}^{\prime}+e(r+1)e\phi^{\prime}+e^{2}rA_{1})',
\label{EXAO}
\end{equation}
and it leads us to evaluate the reduced Hamiltonian density by
plugging in Eqn. (\ref{EXAO}) in the Eqn. (\ref{CHAM1})
\begin{eqnarray}
{\cal H} &=& \frac{\pi_{1}^{2}}{2}
+\frac{1}{2e^2(a-r^2)}(\pi'_1+e(r+1)e\phi^{\prime}+e^{2}rA_{1})^2 +\phi'^{2}\nonumber\\
&+&\pi_{1}A_{0}^{\prime} + e\phi'- e(1+r)\phi' A_1
+\frac{e^2(a+1)}{2} A_1^2. \label{RHAM}
\end{eqnarray}
However, Poisson brackets needs to be replaced by Dirac brackets
\cite{DIR} since the weak condition (\ref{CON2}) is used as strong
condition in {\ref{RHAM}}. The non-vanishing Dirac bracelets for
the fields describing the reduced Hamiltonian are:
\begin{equation}
[A_1(x),\pi_1(y)]^{*}=\delta(x-y),~[\phi(x),
\phi(y)]^{*}=-\frac{\epsilon(x-y)}{4}. \label{DIR}
\end{equation}
The Dirac brackets are found {\it non-canonical}. The reduced
Hamiltonian density (\ref{RHAM}) along with the Dirac brackets
leads to the following coupled equations of motions.
\begin{equation}
\dot{A}_1 = \pi_1- \frac{r}{a-r^2}A_1' -
\frac{1}{e^2(a-r^2)}\pi_1''-\frac{1+r}{e(a-r^2)}\phi'',
\label{EQMA}
\end{equation}
\begin{equation} \dot{\pi}_1 =
ae^2\frac{r^2-a-1}{a-r^2}A_1
 - \frac{r}{a-r^2}\pi_1', \label{EQMP}
\end{equation}
\begin{eqnarray}
\dot{\phi} &=& (1+\frac{(r+1)^2}{2(a-r^2)})\phi'+
e\frac{(r+1)(a+r-r^2)}{a-r^2}A_1\nonumber \\
&-& \frac{1}{2}\frac{1+r}{e(a-r^2)}\pi_1' .\label{EQPH}
\end{eqnarray}
The determination of theoretical spectra needs the decoupling of
the above set of equations (\ref{EQMA}),(\ref{EQMP}) and
(\ref{EQPH}). We observe that for the choice $r=\pm 1$, these
equations leads to Lorentz invariant theoretical spectra that
contains a massive boson with mass $m= \sqrt{\frac{a^2e^2}{a-1}}$
accompanied with {\it a massless chiral boson} having left/right
chirality for $r=\mp 1$. It shows an exact agreement with the
result reported in \cite{KH}. The massless chiral boson can be
thought of in terms of a chiral fermion in the $(1+1)$ dimension.
Thus the presence of chiral fermion ensures that the fermion did
not confine for $r= \pm 1$. We should mention that Lagrangian
density for $r=-1$ can be obtained imposing the constraint
$\tilde{\Omega}_{Ch}=\pi_\phi+\phi'=0$ in the phase-space of the
theory described by Eqn. (\ref{LAGB}) and in this situation the
resulting lagrangian density reds
\begin{eqnarray}
L_{CB} &=& \dot\phi\phi' -\phi'^2 +(1-r)(A_0-A_1)\phi' +
\frac{ae^{2}}{2}(A_0^2 -A_1^2) \nonumber \\
&-&\frac{1}{2}e^2(A_1+rA_0)^2 + \frac{1}{2}(\dot{A_1}-A_0')^2.
\label{LCBMM} \end{eqnarray}
 Comparing Eqn. (\ref{EQMA}) with the
Eqn. $(36)$ of the article \cite{SY}  absence of the term
$-\frac{1+r}{e(a-r^2)}\phi''$ from Eqn. $(36)$ has been found. In
addition to that, equation of motion corresponding to $\phi$, e.g.
$[\phi, H_R]$ was not taken into consideration to come to the
conclusion concerning the theoretical spectra. The model
(\ref{LAGCH}) is also found exactly solvable for $r=0$ and $r >>
1$. Note that the lagrangian (\ref{LAGB}) lands into the
celebrated vector Schwinger model with coupling strength $e$, for
$r=0$ while for $r >> 1$ it can approximately be considered as
Schwinger model described with axial vector interaction with
coupling strength $re$. The models (\ref{LAGB}) however exhibit
exact solvability for $r=0$ and $r >> 1$ with two specific
regularization. The choice $a=0$ and $a=r^2$ works for $r=0$, and
$r >> 1$ respectively. In the second care, it can be thought that
the coupling constant gets scaled by a factor $r$, and the vector
interaction ceases to zero. For these two cases, if chiral
constraint $\Omega_{Ch}=\pi_\phi-\phi'=0$ is imposed in
(\ref{LAGB}), the resulting model describes confinement of fermion
of a definite chirality, and the theoretical spectrum contains
only a missive boson with mass $2e$ and $2re$ for $r=0$ and $r >>
1$ respectively \cite{ARMPLA}. So for the set of values $r=0$,
$r=\pm 1$, and $r
>> 1$ although the model (\ref{LAGB}) is found exactly solvable,
for a generic $r$  after the imposition of chiral constraint the
issue of exact solvability along with confinement aspect of
fermion  remained un settled.

Let us now point out the inconsistencies in the Lagrangian
formulation.  Here the author introduced two ansatzes: Eqn.
$(15)$, and $(16)$ for the fields $\phi$ and $A_\mu$ respectively.
However, none of these two satisfy the Eqns. $(12)/(17)$, $(13)$,
and $(14)$. it is surprising tat Eqn. $(16)$ is designed in an
ambiguous manner to satisfy Eqn. $(18)$ without paying heed to the
Eqn. $(12)/(17)$, $(13)$, and $(14)$ and ambiguously Eqn. $(18)$
is identified as a massive boson!

Therefore, it transpires that from the beginning the author
spoiled both Hamiltonian and Lagrangian formulation in \cite{SY}.
We are afraid that it is irremediable for generic $r$. Albeit, it
is known that the GCSM model (\ref{LAGB}) remains exactly solvable
for a generic $r$, \cite{BASS, MIA, SAR}, but the question of
solvability and  physical sensibility of (\ref{LAGCH}) stood open
for the generic $r$. Our investigation reveals clearly that exact
solvability are manifested only for $r=0$, $r=\pm 1$, and $r>>1$,
subject to the chiral constraint that fits with these permissible
values of $r$ which ensures physical sensibility and confinement
of fermion takes place only for $r=0$ and $r>>1$ (for very large
$r$).

To summarize we reiterate that only the values $r=0, \pm 1$, and
$r>>1$ the model (\ref{LAGB}) may lead to physically sensible
results  after the imposition of chiral constraint suitable for
the values of $r$ that renders exact solvability. However
confinement of fermion occurs only for $r=0$ and $r>>1$. The set
of values $r=\pm 1$ although renders exact solvability, fermion of
a left chirality for $r=-1$ and a fermion of of right chirality
for  $r=+1$  remain unconfined. We would also like to add that the
attempt to establish the physical Lorentz invariance in \cite{SY}
for generic $r$ through Poincare algebra is not trustworthy which
could be understood from the work previous work \cite{SAR}.
However, for the set of values values $r=0$, $r=\pm 1$, and
$r>>1$,  physical Lorentz invariance does  maintained, although
the Lagrangian densities does not have manifestly Lorentz
covariant structure.

\end{document}